\documentclass[preprintnumbers,amsmath,amssymb,twocolumn,pra,showpacs]{revtex4-1}
\usepackage{graphicx,comment}% Include figure files
\usepackage{epstopdf}
\usepackage{dcolumn}% Align table columns on decimal point
\usepackage{bm}% bold math
\usepackage[T1]{fontenc}
\usepackage[latin1]{inputenc}
\usepackage{enumerate}
\usepackage{fancyhdr,color}
\usepackage{footnote}
\usepackage{natbib}

\def\cE{\mathcal{E}}

\def\cO{\mathcal{O}}

\def\Pr{\mathrm{Pr}}

\definecolor{Purple}{rgb}{.7,0.,1.0}

\begin{document}

%\preprint{APS/123-QED}

\title{Security of Decoy-State Protocols for General Photon-Number-Splitting Attacks}

\author{Rolando D. Somma}
\email{somma@lanl.gov}

%\affiliation{
%Los Alamos National Laboratory \\
%Los Alamos, New Mexico 87545 \\
%}

\author{Richard J. Hughes}
\email{rxh@lanl.gov}

\affiliation{
Los Alamos National Laboratory, \\
Los Alamos, New Mexico 87545, USA \\
}

\begin{abstract}
Decoy-state protocols provide a  way to defeat photon-number splitting attacks
in quantum cryptography implemented with weak coherent pulses. We  point out that previous security analyses
of such protocols relied on assumptions about  eavesdropping attacks that considered treating
each pulse equally and independently. We  give an example to demonstrate that,
without such assumptions, the security parameters of previous decoy-state implementations
could be worse than the ones claimed. Next  we consider  more general 
photon-number splitting attacks, which correlate different pulses,
and give an estimation procedure for the number of single photon signals
with rigorous security statements. The impact of our result is that previous analyses of the number of times a decoy-state quantum cryptographic system can be reused before it makes a weak key must be revised.
\end{abstract}

LA-UR 13-20168

\date{\today}% It is always \today, today,
           
             
\maketitle

\section{Introduction: photon number splitting attacks and decoy-state protocols} 
Quantum key distribution (QKD)~\cite{Wie83,BB84,Eke91,Ben92} allows two parties, Alice and Bob, to establish
a common and secret key $S$ that is informationally secure; see~\cite{May01,SP00,Ren05}
and references therein. A widely used setup for QKD is the one 
suggested by Bennett and Brassard (BB84)~\cite{BB84}.  BB84  is ideally implemented by
preparing and  transmitting
single-photon pulses. Information can be encoded in the state of one of two conjugate
polarization bases,
e.g. vertical/horizontal or diagonal/antidiagonal. Only those {photons} 
that were prepared by Alice and detected by Bob in the same basis are useful to build
a sifted key, which forms $S$ after additional steps of information reconciliation and privacy amplification.
Security  follows from the inability of {faithfully} copying quantum information~\cite{Zur82}
and the unavoidable information-disturbance trade-off in quantum mechanics.
Nevertheless, realistic implementations of BB84 use weak coherent photon pulses
that could involve many photons, avoiding the assumptions made in security analyses
~\cite{BBBSS92,SWF07,RPH09}.
Such pulses  could be exploited by Eve, the eavesdropper, to gain access to the (insecure)
distributed key using a so-called photon-number splitting (PNS) attack~\cite{Lut00,LJ02}.
In a simple proposed PNS attack, Eve measures the number of photons in the pulse, $n$.
If $n=1$, Eve blocks the pulse.
If $n \ge 2$, Eve ``splits'' the pulse to obtain a copy of a single photon with the correct
polarization and keeps it in
her quantum memory.   Eve could then obtain a full copy of $S$ by making measurements 
of her photons in the correct polarization bases, which are known after a public discussion between Alice and Bob. 
Since Alice and Bob cannot measure $n$, a PNS attack may go undetected.
Our goal is to provide a protocol for secure QKD
in the presence of PNS attacks.

A simple approach to overcome a PNS attack
considers reducing the probability
of multi-photon pulses by reducing the coherent-pulse
intensities. The  drawback with this approach
is that the probability of creating single-photon pulses
is also reduced. Then, the rate at which bits to build $S$
are sifted is far from optimal~\cite{LJ02,GLLP04}.
Another approach is to use decoy states, that allow
to detect PNS attacks without a substantial
reduction on the rate of sifted bits if Eve is not present~\cite{Hwa03,LMC05,RH09}. 
In a decoy-state protocol (DSP), one of several 
weak coherent sources is randomly selected for each pulse.
Such sources create pulses of different intensities (mean photon numbers).
This gives Alice and Bob  a means to estimate $f_0$ and $f_1$, the number
of Bob's detections due to empty and single-photon pulses prepared by Alice, in the same basis, respectively.
The values of $f_0$ and $f_1$ are important to determine $|S|$, the length of the secure key.
For example, in the discussed PNS attack, $f_1$ is substantially smaller  than its 
value when Eve is not present, and so is $|S|$.

In more detail, we let $K \gg 1$ be the total number of pulses
prepared by Alice. We first assume that the channel is non-adversarial,
i.e.  no eavesdropping attacks
are present.
If the pulse has a random phase,   
the number of photons  it contains is
sampled according to the Poisson distribution: 
\begin{align}
\label{eq:Poisson}
p_n^\mu=\Pr(n|\mu) = e^{-\mu}\frac{ \mu^n}{n!} \; ,
\end{align}
where $\mu$ is the mean photon number
for that source and $\mu \le 1 $ in applications.  We let $\eta$ be the transmission/detection efficiency
of the quantum channel shared by Alice and Bob. If $b=1$ ($b=0$) denotes the event in which Bob
detects a non-empty (empty or vacuum) pulse,
\begin{align}
\nonumber
y_n = \Pr(b=1|n) 
\end{align}
is the probability of a detection by Bob given that Alice's prepared pulse contained $n$ photons.
$y_n$ is the so-called $n$-photon yield and
$y_n < 1$ due to losses in the channel.
For $n \ge 1$, we may assume
\begin{align}
\nonumber
y_n =1 - (1-\eta)^n \; ,
\end{align}
which is a good approximation in applications. For $n=0$, $y_0>0$ denotes Bob's detector dark-count rate. 
The total probability of  Bob detecting a pulse (in any one cycle) is the total yield
\begin{align}
\label{eq:totalyield}
Y(\mu) &= \Pr(b=1) \\
\nonumber &= \sum_{n \ge 0} \Pr(n|\mu) y_n \\
\nonumber
& = e^{-\mu} y_0 + 1 - e^{-\mu \eta} \; .
\end{align}
$Y(\mu)$ can be estimated by Alice and Bob, via public discussion, from the frequency of detections
after all pulses were transmitted.

In QKD, we allow Eve to manipulate the parameters that characterize the channel
at her will. We use the superscript $\cE$ to represent the interaction of Eve with
the communication. For example, $y^\cE_n$ denotes the $n$-photon
yield  in the presence of Eve. In a general intercept-resend attack, Eve
may intercept a pulse and resend a different one. That is, each detection
by Bob is not guaranteed to come from the same pulse that Alice prepared.
In a simple PNS attack, Eve makes non-demolition measurements of $n$.
With this information, Eve sets $y^\cE_1=0 \ne y_1$ and
$y^\cE_n \ge y_n$ for $n \ge 2$, so that
\begin{align}
\nonumber
Y(\mu) \approx Y^\cE(\mu) \; .
\end{align}
Then, if Alice and Bob can only estimate the total yields,
a PNS attack could be ``invisible'' with the right choices of $y_2^\cE,y_3^\cE,\ldots$.
To increase the multi-photon yield,
Eve may use an ideal channel to resend the pulses. 
(Note that sophisticated 
PNS attacks that do not change the Poisson distribution are possible~\cite{LJ02}.)
A PNS attack allows Eve to have the full key $S$
if 
\begin{align}
\nonumber
\Pr(n \ge 2 | \mu) \ge Y(\mu) \; .
\end{align}
In this case, Eve possesses a photon
with the same polarization as that of the pulse detected by Bob
and no single-photon pulses are involved in creating $S$.
Only if $\Pr(n \ge 2 | \mu) < Y(\mu)$ some security guarantees are possible~\cite{GLLP04}. 
Such an inequality is satisfied when $\mu \approx \eta$, implying a rate for sifted bits
of order $\eta^2$ [Eq.~\eqref{eq:totalyield}]. This is undesirably
small ($\eta \ll 1$).  

\begin{figure}[ht]
  \centering
  \includegraphics[width=7.5cm]{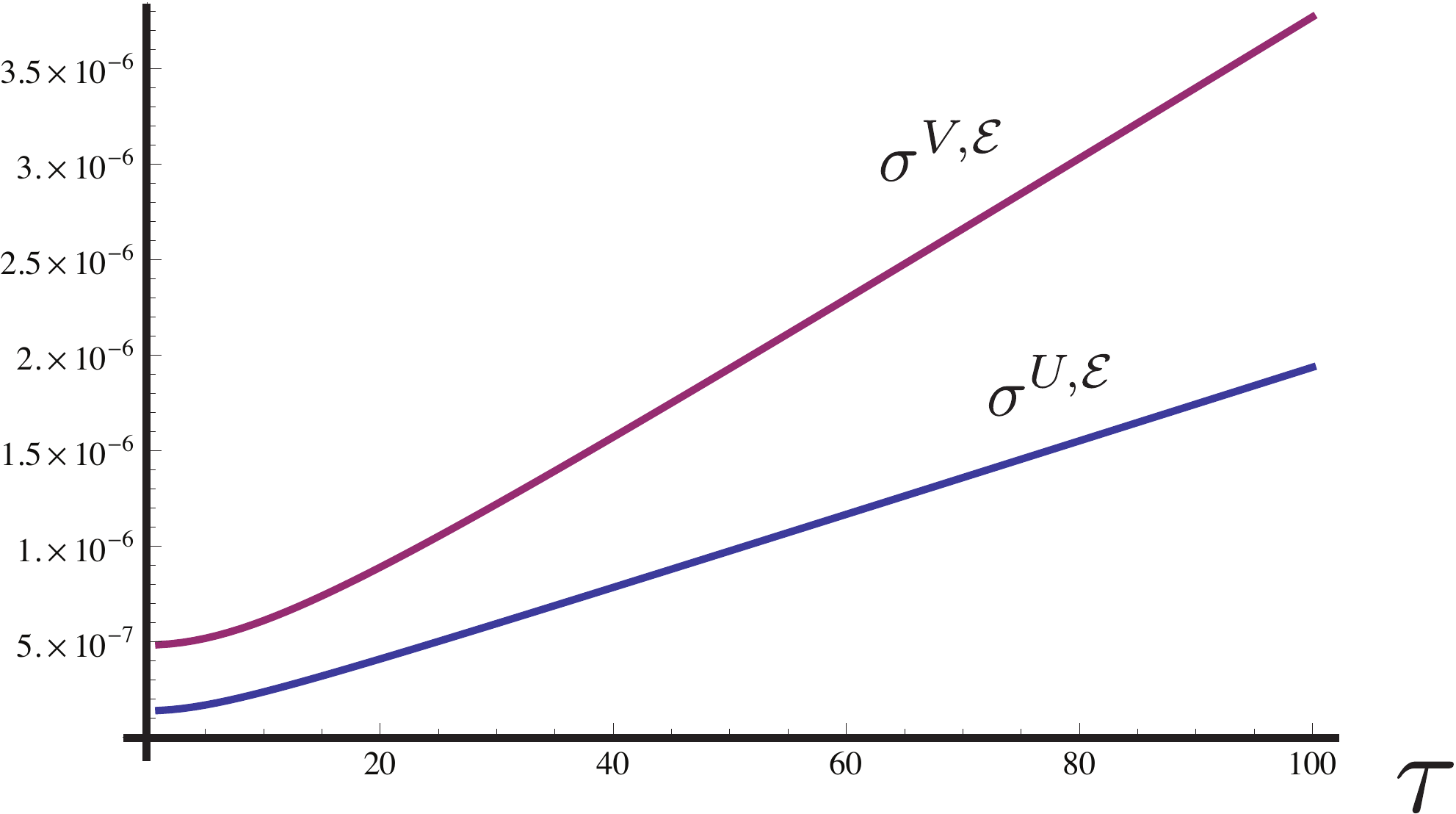}
  \caption{The standard deviations $\sigma^{U,\cE}$ and $\sigma^{V,\cE}$   for an attack
  in which Eve correlates $n$-photon pulses according to the value
  of   $\tau$. The channel parameters are $1 \le \tau \le 100$, $K=10^{10}$, $q^U=0.01$, $q^V=0.0275$, $\mu^U=0$, $\mu^V=0.063$,
  $\eta=10^{-3}$, and $y_0=2.10^{-6}$~\cite{RH09}.  The results in Sec.~\ref{sec:decoys} are recovered for $\tau=1$.}
  \label{fig:Attack1}
\end{figure}

Remarkably, DSPs give an optimal rate of order $\eta$ with
small resource overheads.   
A goal in a DSP is to estimate $y^\cE_0$ and $y^\cE_1$, 
which provide a lower bound on $f_0^\cE$ and $f_1^\cE$,
respectively. Empty and single-photon pulses cannot be split and the information 
carried in their polarization cannot be faithfully copied,
making them useful to create a secure key. For the estimation,
Alice uses photon sources with different values of $\mu$, but are identical otherwise.
{In a conventional DSP, it is assumed that Eve's PNS attack treats every $n$-photon pulse
equally and independently, regardless of its source. That is, Eve's attack
is simulated by independent and identically distributed (i.i.d.) random variables.
The total yield in this case is, for any given $\mu$,
\begin{align}
\label{eq:Evetotalyield}
Y^\cE(\mu) = \sum_{n \ge 0} p_n^\mu \ y^\cE_n \; .
\end{align}
Equation~\eqref{eq:Evetotalyield} describes mathematically what we denote as the i.i.d. assumption.
 It follows that
\begin{align}
\nonumber
y^\cE_0 &= Y^\cE(\mu) |_{\mu=0} \; , \\
\label{eq:singlephotonTE}
y^\cE_1 &= \partial_\mu \left[ e^\mu Y^\cE(\mu) \right] |_{\mu=0} \; .
\end{align}
Then, if Eve's attack satisfies $Y(\mu) \approx Y^\cE(\mu)$ for all $\mu$, 
\begin{align}
\nonumber
 y^\cE_0 \approx y_0 \; , \; y^\cE_1 \approx y_1 = \eta \; .
\end{align}
That is, by being able to estimate $Y^\cE(\mu)$ for two values of $\mu \ll 1$ via public discussion, 
 Alice and Bob can restrict Eve's attack so that the dark-count rate
 and single-photon yield
 are almost unchanged {from the non-adversarial case. In addition, if a third source with 
 $\mu \approx 1$ is randomly invoked,
an optimal key rate of order $\eta$ will be achieved. 

In reality, the estimation of $y_0^\cE$ and $y_1^\cE$ is subject to finite statistics and
can be technically involved. {Nevertheless, the i.i.d. assumption in Eq.~\eqref{eq:Evetotalyield}
allows Alice
and Bob to gain information about Eve's attack by running the
 protocol and analyzing the (binomial) distributions of
the detection events for each source. However, we remark that if Eve were to correlate her attacks,
the i.i.d. assumption and the corresponding security analyses would be invalid.
This is the main motivation behind our analysis.
}

In this paper, we  give an example that shows
how the i.i.d. assumption can be simply bypassed by Eve,
resulting in security parameters that are worse from those obtained 
under  the  assumption. We then analyze the security of DSPs
for general {PNS} attacks.
 Our main result is
 an estimation procedure that gives 
a lower bound on $f_0^\cE$ and $f_1^\cE$,
with a confidence level that is  an input
to the estimation procedure. Our security analysis
does not use the i.i.d. assumption and is particularly relevant
when Eve performs a PNS attack that could correlate different pulses in one session or even
different  sessions. We 
compare some results obtained by our estimation procedure with those obtained by
using the i.i.d. assumption, and emphasize the important of our procedure.

%%%%%%%%%%%%%%%%%%%%%%%%%%%%%%%%%%%%%%%%%%%%%%%
\section{The security parameter, the i.i.d. assumption, and finite statistics}
%%%%%%%%%%%%%%%%%%%%%%%%%%%%%%%%%%%%%%%%%%%%%%%
\label{sec:decoys}

Of high significance in cryptographic protocols is $\epsilon$, 
the so-called security parameter. $\epsilon$ measures
the deviation of a real protocol implementation from an ideal one.
We use the same definition used in Ref.~\cite{Ren05}, that states that a real QKD
protocol is $\epsilon$-secure if it is $\epsilon$-indistinguishable
from a perfectly secure and ideal one.  This definition is equivalent
to a statement on the trace norm of the difference between
the quantum states resulting from the real and ideal protocol, respectively.
It implies that a QKD protocol that is $\epsilon$-secure can be safely reused
order $1/\epsilon$ times without compromising its security.

 Usually, one fixes a value for $\epsilon$  and then determines the size
 of $S$ based on several protocol performance parameters.  These parameters include
 the number of pulses sent by Alice, the number of pulses detected by Bob,
  and the estimated bit error rates at each mean photon number. 
 For DSPs, $\epsilon$ has a component  $\epsilon_{\rm DSP}$
 that determines the confidence
 level in the estimation of  a lower bound of $f_0^\cE$ and $f_1^\cE$, due to finite statistics.

A possible way to obtain such lower bounds, under the i.i.d. assumption, 
is the one followed in Ref~\cite{RH09}.
In this case, we consider a DSP with three sources, $i=U , V ,W$.
The mean photon number  in each pulse, for each source, 
is $\mu^U=0$, $\mu^V \ll 1$,  and $\mu^W \in \cO(1)$.
Each source $i$ randomly prepares a pulse with probability $q^i$ and
we let {$K^i$} be
the total number of pulses for that source. {$K^i$}
is known to Alice and Bob by public discussion after all pulses are sent
 and {$K^i \approx q_i K$
when $K \gg 1$.}
We write $D^{i,\cE}$ for the random variable
that counts the number of pulses from source $i$ detected by Bob
under the presence of Eve~\cite{Note3}.
The  exact value that $D^{i,\cE}$
takes in a session can also be obtained by Alice and Bob
via public discussion after the pulses were transmitted.

Under the i.i.d. assumption [Eq.~\eqref{eq:Evetotalyield}], $D^{i,\cE}$
is sampled according to the binomial distribution. Then, {$D^{i,\cE}/K^i$
is an estimator of the total yield $Y^\cE(\mu^i)=E[D^{i,\cE}/K^i]$, where $E[.]$
denotes the mean value. That is, for a given $\bar \epsilon_{\rm DSP}$,}
we can establish confidence intervals 
\begin{align}
\label{eq:DSPbinomial}
\frac{D^{i,\cE}}{K^i} + c(\bar \epsilon_{\rm DSP}) \sigma^{i,\cE} \ge Y^\cE(\mu^i) \ge \frac{D^{i,\cE}}{K^i} - c(\bar \epsilon_{\rm DSP}) \sigma^{i,\cE} \; ,
\end{align}
with confidence level $1-\bar \epsilon_{\rm DSP}$. The constant $c$ depends on $\bar \epsilon_{\rm DSP}$ 
 and can be obtained using Chernoff's bound~\cite{Ho63} -- see Appendix~\ref{app:chernoff}. 
The standard deviation in this case is
\begin{align}
\label{eq:binvariance}
\sigma^{i,\cE} \approx \sqrt{\frac {Y^{\cE}(\mu^i) (1-Y^{\cE}(\mu^i))}{q^i {K}} } \; .
\end{align}
Using Eq.~\eqref{eq:Evetotalyield} for $Y^{\cE}(\mu^i)$,
we can search for the minimum values of $y_0^\cE$ and $y_1^\cE$
that satisfy Eqs.~\eqref{eq:DSPbinomial}, e.g. by executing a linear program.
Both $y_0^\cE$ and $y_1^\cE$ can then be used to obtain the desired lower
bounds on $f_0^\cE$ and $f_1^\cE$, respectively, with  corresponding confidence
level $1-\epsilon_{\rm DSP}$. This last step also requires using the i.i.d. assumption.

We remark that Eq.~\eqref{eq:DSPbinomial} does not properly regard
the problem of inferring a distribution for $Y^\cE(\mu^i)$ from the known
 $\frac{D^{i,\cE}}{K^i}$, a problem that would require knowledge on the prior
 distribution of $Y^\cE(\mu^i)$.

%%%%%%%%%%%%%%%%%%%%%%%%%%%%%%%%%
%%%%%%%%%%%%%%%%%%%%%%%%%%%%%%%%%%%%
\section{Increasing the length of confidence intervals: An attack}
\label{sec:attack}
 The  analysis in Sec.~\ref{sec:decoys} used the i.i.d. assumption that resulted in a 
 value for $\sigma^{i,\cE}$  given by Eq.~\eqref{eq:binvariance}.
 Nevertheless, the actual value of $\sigma^{i,\cE}$ could be much higher
 in more general PNS attacks. For the same confidence level,
  a bigger $\sigma^{i,\cE}$ implies  a ``wider'' confidence interval for the estimation of 
  the yield $Y^\cE(\mu^i)$ (Appendix~\ref{app:chernoff}),
 and thus smaller lower bounds on $f_0^\cE$ and $f_1^\cE$. 
 The overall result
 in the DSP is a secret key $S$ of smaller size for the same security parameter.
 
To illustrate how Eve can bypass the i.i.d. assumption, 
we suggest a potential attack that results in almost no change for
the total yields (i.e., $Y(\mu^i) \approx Y^{\cE}(\mu^i)$)~\cite{Note1} but the variances $\sigma^{i,\cE}$
are increased 
with respect to those of the binomial distribution [Eq.~\eqref{eq:binvariance}].
The suggested attack could be detected by Alice and Bob by estimating the variances
directly via public discussion. Nevertheless, it still shows that a better analysis of the security of DSPs
 is needed to make rigorous claims.

In the attack,
Eve first picks an integer value for $\tau \ge 1$,
where $\tau^2$ denotes a scale for a variance or ``correlation'' of a particular distribution.
Eve receives all pulses from Alice and we let $k_n$ be the total number
of $n$-photon pulses in the protocol. Note that the exact value of $k_n$ is known to Eve but not  to
Alice and Bob.  In general, $k_n$ is sampled according to the binomial distribution
\begin{align}
\nonumber
\Pr(k_n) =  \begin{pmatrix}  K \cr k_n \end{pmatrix} (p_n)^{k_n} (1-p_n)^{K-k_n} \; ,
\end{align}
where $p_n$ is the probability of a pulse containing $n$ photons: $p_n = \sum_i q^i p_n^{\mu^i}$.
The mean and variance for such distribution are 
\begin{align}
\nonumber
E[k_n] &=p_n K \; , \\
\nonumber
\sigma^2_{k_n} &= p_n (1-p_n) K \; .
\end{align}

Given $k_n$, Eve randomly picks a value for $d^{\cE}_n \in \{0,1\ldots,k_n\}$, 
where $d_n^\cE=\sum_i d_n^{i,\cE}$ is the total number of detections due to $n$-photon pulses prepared by Alice.
In particular, we assume that Eve can control $d^{\cE}_0$, which determines the dark-count rate. 
The distribution
associated with $d^{\cE}_n$ has the following properties: 
\begin{align}
\label{eq:dnDIST}
E[d^{\cE}_n|k_n] &= y_n k_n \; , \\
\nonumber
\sigma^2_{d^{\cE}_n|k_n} & = \tau^2 y_n (1-y_n)  k_n \; .
\end{align}
We let $d_n^{i,\cE}$ } be the number of $n$-photon pulses, prepared by Alice's  $i$th source only, and detected by Bob.
The exact value of $d_n^{i,\cE}$ is unknown to all parties. Because Eve does not know the source being used
in the DSP, $d_n^{i,\cE}$ is sampled according to the binomial distribution when given $d_n^{\cE}$: 
\begin{align}
\label{eq:dnidist}
\Pr(d^{i,\cE}_n | d^\cE_n) = \begin{pmatrix} d^\cE_n \cr d_n^{i,\cE} \end{pmatrix} (q^i_n)^{d_n^{i,\cE}} (1-q^i_n)^{d^\cE_n-d_n^{i,\cE}} \; ,
\end{align}
 where 
\begin{align}
q_n^i = \frac {q^i e^{-\mu^i} (\mu^i)^n}{\sum_{i'=U,V,W} q^{i'} e^{-\mu^{i'}} (\mu^{i'})^n} \; .
\end{align}
The distribution associated with $d_n^{i,\cE}$ satisfies
\begin{align}
\nonumber
E[d_n^{i,\cE}|d_n^{\cE}] &= q_n^i d_n^{\cE} \; , \\
\nonumber
\sigma^2_{d_n^{i,\cE}|d_n^\cE} & = q_n^i (1-q_n^i) q_n^\cE \; .
\end{align}

As in Sec.~\ref{sec:decoys}, we let $(\sigma^{i,\cE})^2$ be the variance
associated with the random variable $Z^{i,\cE} =D^{i,\cE}/K^i$, where 
\begin{align}
\label{eq:constrains}
D^{i,\cE} = \sum_{n \ge 0} d_n^{i,\cE} \; , 
\end{align} 
and $E[Z^{i,\cE}]=Y^\cE(\mu^i)$.
An accurate estimate of $Z^{i,\cE}$ can be obtained
if we approximate $K^i \approx   q^i K $, in the limit of large $K$.
In addition, because $K$ is fixed, the variables $k_n$ are not independent.
However, in the large-$K$ limit, $k_n$ can also be approximated by its mean value.
It implies that the $k_n$ are almost independent and so are the $d_n^{\cE}$
and $d_n^{i,\cE}$ for different values of $n$. Under these approximations, 
\begin{align}
\label{eq:dnvariance}
(\sigma^{i,\cE})^2 \approx \frac 1 {( q^i K )^2} \sum_{n \ge 0} \sigma^2_{d_n^{i,\cE}} \; .
\end{align} 
In Appendix~\ref{App:AttackStatistics} we show that 
\begin{align}
\label{eq:dnivariance}
\sigma^2_{d_n^{i,\cE}} =[ (\tau^2-1) q_n^i (1-y_n) + (1-q_n^i y_n p_n)] q_n^i y_n p_n K  \; .
\end{align} 
By inserting} Eq.~\eqref{eq:dnivariance} in Eq.~\eqref{eq:dnvariance}, we can obtain the variances as a function of $\tau$.
In Fig.~\ref{fig:Attack1} we compute $\sigma^{U,\cE}$ and $\sigma^{V,\cE}$.
 The i.i.d. assumption discussed in Sec.~\ref{sec:decoys}
corresponds to $\tau=1$ -- see Appendix~\ref{App:AttackStatistics}. Using these results in Eq.~\eqref{eq:DSPbinomial} yields  
wider confidence intervals 
for the same confidence level.

%The confidence intervals for the number of detections
%due to empty and single-photon pulses strongly depend on the attack
%conducted by Eve. Such confidence intervals could be much
%wider than the ones obtained by assuming that the attack
%treats every pulse equally and independently (the i.i.d. assumption). \rol{ This follows from
%the Chernoff bound~\cite{Ho63},  that results in a similar value of
%$c(\bar \epsilon_{\rm DSP})$ for the binomial distribution discussed in Sec.~\ref{sec:decoys} [Eq.~\eqref{eq:DSPbinomial}]
%and for the distribution of $Z^{i,\cE}$ resulting from the attack suggested in this section -- see Appendix~\ref{app:chernoff},
%Eq.~\eqref{eq:chernofferror0}.

To  illustrate our point further, we consider a simple attack in which a single
source $U$ is used to estimate the dark-count rate. Here, $\mu^U=0$
and $d_0^\cE=D^{U,\cE}$ is known. In the non-adversarial case, $d_0^\cE$ is sampled according to the binomial distribution
with probability $y_0$ and  known sample size $k_0=K^U$. Nevertheless, for the correlated attack, we assume that
Eve ``receives'' the $K^U$ pulses and groups them according to blocks of size $\tau^2$. Then,
Eve will force (prevent) the detection of all pulses in any one block with probability $y_0$ ($1-y_0$).
The random variable $d_0^\cE$ for the correlated attack satisfies
\begin{align}
\nonumber
E[d_0^\cE] & = y_0 k_0 \; , \\
\nonumber
\sigma^2_{d_0^\cE} (\tau)& =  [y_0 (\tau^2)^2 - (y_0 \tau^2)^2]\frac{k_0}{\tau^2} =  \tau^2 y_0 (1-y_0) k_0 \; ,
\end{align}
and $\tau=1$ corresponds again to the i.i.d. assumption [see Eq.~\eqref{eq:dnDIST}].  

\begin{figure}[ht]
  \centering
  \includegraphics[width=7.8cm]{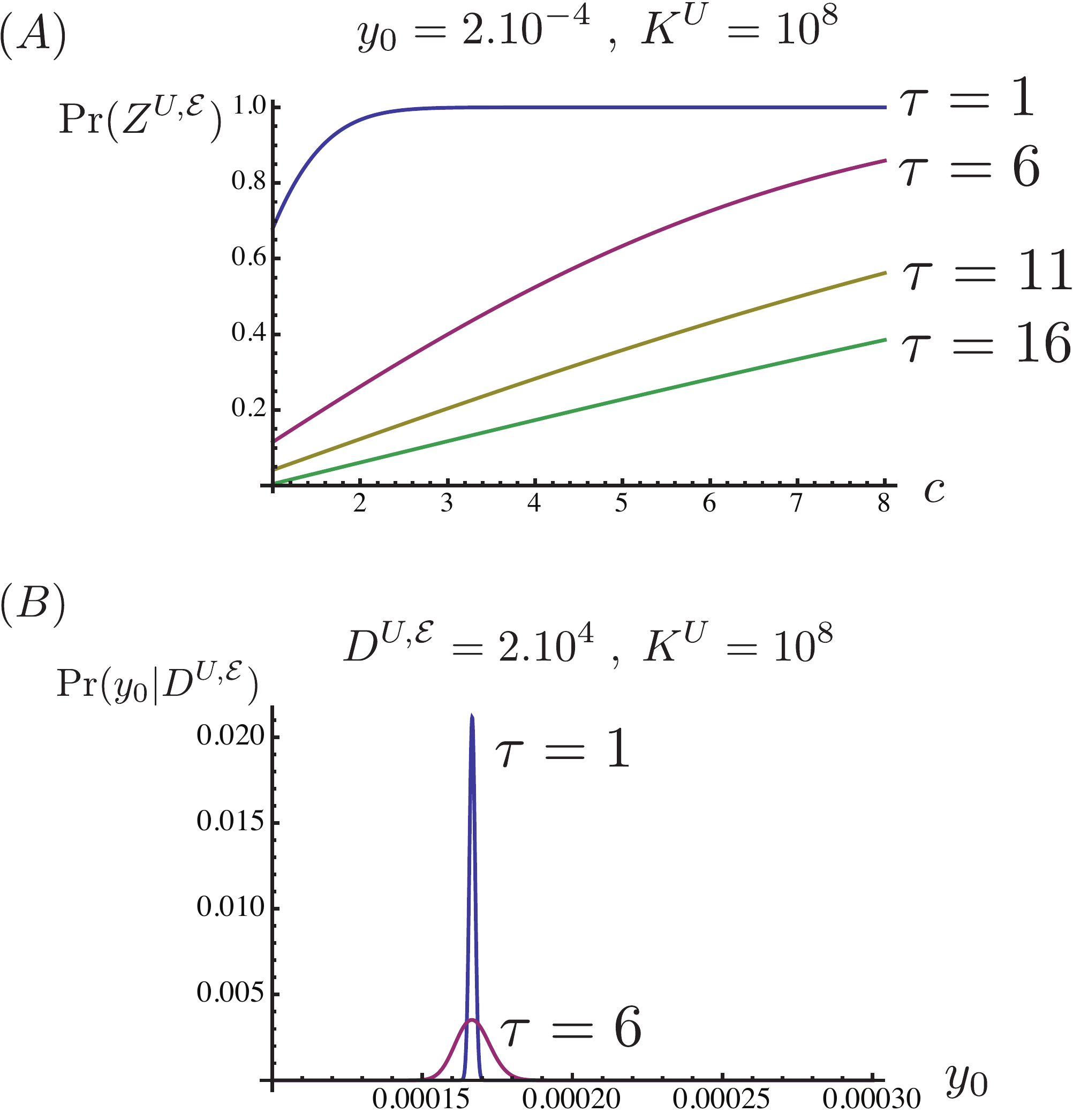}
  \caption{Estimation of dark counts. (A) Confidence intervals for different correlated attacks,
  parametrized by $\tau$, and confidence bounds, parametrized by $c$. (B) Bayesian estimation of $y_0$, the mean dark-count rate,
  assuming a uniform prior and for different correlated attacks [Eq.~\eqref{eq:BayesEst}].}
  \label{fig:DarkCount}
\end{figure}

In Fig.~\ref{fig:DarkCount} (A), 
we plot the probability that $Z^{U,\cE}=D^{U,\cE}/K^U$ satisfies
\begin{align}
\nonumber
E[Z^{U,\cE}] + c \sigma_{d_0^\cE} (1) \ge  Z^{U,\cE} \ge E[Z^{U,\cE}] - c \sigma_{d_0^\cE} (1) \; , 
\end{align}
for different values of $c$ and $\tau$. For $\tau=1$, such a probability
corresponds to the confidence level in Eq.~\eqref{eq:DSPbinomial}. $E[Z^{U,\cE}]=y_0$ in this example.
For the inverse problem, namely the estimation of $y_0$
from $D^{U,\cE}$ and $K^U$, Eq.~\eqref{eq:DSPbinomial} may be incorrect.
We may then assume a uniform prior distribution for $y_0 \in [0,1]$,
and obtain the posterior distribution as
\begin{align}
\label{eq:BayesEst}
 \Pr (y_0 | & D^{U,\cE}) = \Pr (D^{U,\cE}|y_0) \Pr(y_0)/\Pr(D^{U,\cE})  \\
\nonumber
& \propto \begin{pmatrix} K^U/\tau^2 \cr D^{U,\cE}/\tau^2 \end{pmatrix} y_0^{D^{U,\cE}/\tau^2}
(1-y_0)^{(K^U-D^{U,\cE})/\tau^2} \; ,
\end{align}
which is plotted in Fig.~\ref{fig:DarkCount} (B). 
Our results demonstrate that, for a fixed security parameter, the accuracy in the estimation of the dark-count rate 
strongly depends on Eve's attack and can be substantially different from the
one obtained under the i.i.d. assumption ($\tau=1$).

%%%%%%%%%%%%%%%%%%%%%%%%%%%%%%%%%%%%%%%%
\section{Security of DSP: Correlated PNS attacks} 
\label{sec:generalsecurity}
We  go beyond the i.i.d. assumption and
study more general and correlated PNS attacks,
in which Eve has  full control
on Bob's detection events.
The secure key-rate in a realistic implementation of QKD is~\cite{RH09} 
\begin{align}
\label{eq:keyrate}
s   \ge f_0^{\cE *} +      f_1^{\cE *} - \kappa_{\rm EC}  F^{\cE} H_2({\rm BER}) -    
 \kappa_{\rm PA} f_1^{\cE *} H_2(b_1^{\max})    \; ,
\end{align} 
which determines the size of the distributed key as $|S|=sK$. 
$F^\cE$ is the total number of pulses detected by Bob
and prepared by Alice in the same basis, that are useful for the sifted key. 
In BB84, $F^\cE \approx D^\cE/2$, where $D^\cE$ is the total number of detections.
$f_n^{\cE *}$ is a lower bound on $f_n^\cE$, the number of $n$-photon pulses
prepared and detected in the same basis.
$H_2(.)$ is the Shannon entropy,
$\kappa_{\rm EC}$ and $\kappa_{\rm PA}$ are coefficients due to the error correction and privacy amplification
steps, BER is the total bit error rate, and $b_1^{\max}$ is an upper 
bound to the bit error rate due to single-photon pulses only.

\begin{figure}[ht]
  \centering
  \includegraphics[width=7.8cm]{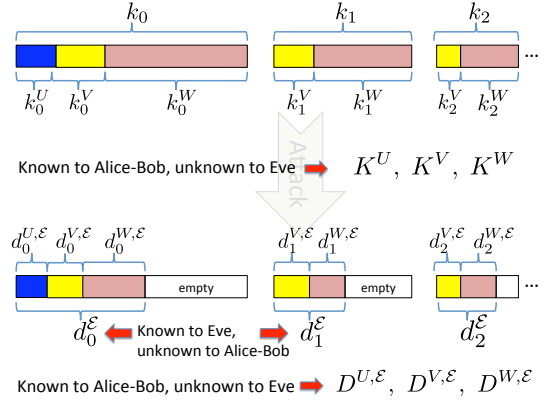}
  \caption{A general PNS attack with three decoy sources, $\mu_U=0$, $\mu_V \ll 1$,
  and $\mu_W \in \cO( 1)$. Each block represents the number of pulses with $n=0,1,2,\ldots$,
  respectively.
  The random variables $k_n$ indicate
  the number of $n$-photon pulses prepared by Alice and the superscript $i$ denotes the source
  used for such pulses.
  Eve's attack  controls the number of detections
  by Bob, due to $n$-photon pulses, through $d_n^{\cE}$.
  }
  \label{fig:InterceptResend1}
\end{figure}

In a DSP, we characterize a general PNS attack 
by the distribution 
\begin{align}
\Pr(d_0^{\cE},d_1^{\cE},\ldots|k_0,k_1,\ldots) \; ;
\end{align} 
See  Fig.~\ref{fig:InterceptResend1} for an example.
Our goal is to build an estimation procedure that places 
confidence intervals on $f_0^\cE =\sum_i f_0^{i,\cE}$ and $f_1^\cE = \sum_i f_1^{i,\cE}$
from the known $D^{i,\cE}$. 
These intervals ultimately imply a lower bound on $s$ --
see Eq.~\eqref{eq:keyrate}.

We  assume that there are three sources satisfying $\mu_U=0 < \mu_V < \mu_W$,
and $\mu_W \in \cO(1)$. Nevertheless, our analysis can be easily generalized 
to the case in which more sources are present, where the estimation is more accurate.
For each source, Bob's detections satisfy Eq.~\eqref{eq:constrains}.
If a simple relationship between each $d_n^{i,\cE}$ and $d_n^{\cE}$  could be found, we could execute 
a program to solve Eqs.~\eqref{eq:constrains}. 
Such a relationship could be obtained from the binomial distribution 
associated with $d_n^{i,\cE}$, when given ${d_n^{\cE}}$ [Eq.~\eqref{eq:dnidist}].

\begin{figure}[ht]
  \centering
  \includegraphics[width=7.8cm]{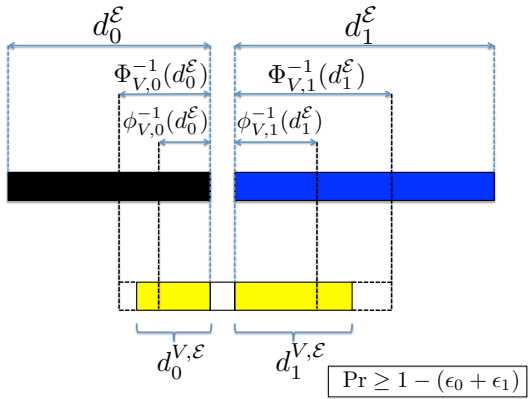}
  \caption{Upper and lower bounds on $d_0^{V,\cE}
 + d_1^{V,\cE} \approx d^{V,\cE}$. The yellow blocks
 represent the number of pulses from source $V$ with $n=0$ and $n=1$.
  The dark and blue blocks represent the total number of pulses
 with $n=0$ and $n=1$, respectively. The confidence level for this case
  is not smaller than $1-(\epsilon_0 + \epsilon_1)$.}
  \label{fig:BoundsFig}
\end{figure}

Our estimation procedure uses ${d_n^{i,\cE}}$ to determine the confidence
intervals 
\begin{align}
\label{eq:upperlowerbound1}
 \Phi_{i,n} (d_n^{i,\cE}) \ge d_n^{\cE} \ge \phi_{i,n} (d_n^{i,\cE}) \; .
\end{align}  
The corresponding confidence level for each inequality is $1-\epsilon_n/2$. 
The upper and lower bounds
are monotonic and invertible functions.
Then, 
\begin{align}
\label{eq:upperlowerbound2}
& \phi_{i,n}^{-1}(d_n^\cE)  \ge  d_n^{i,\cE} \ge \Phi^{-1}_{i,n}( d_n^\cE) \; ,
\end{align} 
with the same confidence levels. 
Such confidence levels do not result from the binomial
distribution as we are analyzing the inverse problem, namely the estimation of $d_n^\cE$
from the available information (i.e., $D^{i,\cE}$ and $K^i$). 
From Eqs.~\eqref{eq:constrains} and \eqref{eq:upperlowerbound2},
we obtain 
\begin{align}
\label{eq:upperlowerbound3}
& \sum_{n \ge 0} \phi_{i,n}^{-1}(d_n^\cE)  \ge  D^{i,\cE}   \ge \sum_{n \ge 0} \Phi^{-1}_{i,n}( d_n^\cE) \; ;
\end{align} 
See Fig.~\ref{fig:BoundsFig} for an example.

Next, our estimation procedure executes a program
to obtain  $d_0^{\cE *}$ and $d_1^{\cE *}$, the corresponding
smallest values of $d_0^\cE$ and $d_1^\cE$, 
subject to the constraints  given by Eqs.~\eqref{eq:upperlowerbound3}.
From the union bound, the confidence level in such values is $1- \bar \epsilon_{\rm DSP}$, with
\begin{align}
\label{eq:securityparameter0}
\bar \epsilon_{\rm DSP} \le 3 \sum_{n \ge 0} \epsilon_n \;  ,
\end{align}
when three sources are used.
Since $f_0^{\cE*}$ and $f_1^{\cE*}$  are sampled according to a 
binomial distribution when given $F^\cE$ (i.e.,
the preparation and detection basis are random),
we obtain
\begin{align}
\label{eq:desiredbound}
 f_n^{\cE *} = F^\cE \frac{ d_n^{\cE *} }{D^\cE} - c(\bar\delta_{\rm DSP}) \sqrt{F^\cE \frac{ d_n^{\cE *} }{D^\cE}  \left(1-\frac{ d_n^{\cE *} }{D^\cE} \right)} \; ,
\end{align}
where the constant $c( \bar \delta_{\rm DSP}) \ge 0$ can be obtained using  
Eq.~\eqref{eq:chernofferror}. 
The overall confidence level for the key rate $s$
is $1-\epsilon_{\rm DSP}$, where the security
parameter satisfies
\begin{align}
\label{eq:securityparameter}
\epsilon_{\rm DSP} \le \bar \epsilon_{\rm DSP} + \bar \delta_{\rm DSP} \; .
\end{align}
In the next section we obtain the confidence intervals and levels specifically for our method.

\vspace{0.5cm}

%%%%%%%%%%%%%%%%%%%%%%%%%%%%%%%%%%%%
\subsection*{Confidence intervals for the estimation procedure}
\label{sec:confidencebounds}
Our method takes $\epsilon_{\rm DSP}$ as input
and outputs $f_0^{\cE*}$ and $f_1^{\cE*}$. 
To satisfy Eq.~\eqref{eq:securityparameter},  we can  set
\begin{align}
c(\bar \delta_{\rm DSP}) & = 2 \sqrt{|\log (\epsilon_{\rm DSP}/2)|}
\end{align}
and
\begin{align}
\label{eq:nphotonerror}
\epsilon_n &= (\epsilon_{\rm DSP}/12) (1/2)^n \; 
\end{align}
[see Eqs.~\eqref{eq:chernofferror} and \eqref{eq:firsterror}].
Next, we will find $d_0^{\cE*}$ and $d_1^{\cE*}$ as required by Eq.~\eqref{eq:desiredbound}.  

If $\phi$ depends on $d_n^{i,\cE}$ only,  
the probability that $d_n^{\cE}$ is smaller than $\phi$ is
 \begin{align}
\label{eq:errorbound2}
 \sum_{d_n^\cE=0}^{K} \Pr(d_n^\cE) \sum_{d_n^\cE \ge d_n^{i,\cE} > u_n^i} \Pr(d_n^{i,\cE}|d_n^\cE)
 =\frac{\epsilon_n}{2}  \; ,
\end{align} 
with
\begin{align}
\nonumber
u_n^i = \phi_{i,n}^{-1} ({d_n^\cE}) \; .
\end{align}
When given {$d_n^\cE$}, the random variable {$d_n^{i,\cE}$} is sampled according
to Eq.~\eqref{eq:dnidist}.
From Chernoff's bound (Appendix~\ref{app:chernoff})
\begin{align}
\epsilon_n 
\label{eq:Hoeffdingbound}
 { \le 2 \max_{0 \le d_n^\cE \le K} \exp \left \{ - \frac {(u_n^i - q_n^i d_n^\cE )^2} {4 q_n^i (1-q_n^i) d_n^\cE}  \right \} \; , }
\end{align}
and we choose the lower bound so that
\begin{align}
\label{eq:lowerbound}
\phi_{i,n} ({d_n^{i,\cE}}) = { \frac{d_n^{i,\cE}} {q_n^i} - c_n \frac{ 1-q_n^i}{2 q_n^i} \left[ \sqrt{ c_n^2 + \frac{ 4 d_n^{i,\cE}} { (1-q_n^i)^2}} -c_n \right] } \; ,
\end{align}
with $c_n \ge 0$. 
The error probability satisfies
\begin{align}
\label{eq:errorbound3}
\epsilon_n \le  2   \exp \left \{ -    c^2_n  /4  \right \}\; ;
\end{align}
See Appendix~\ref{app:errors}. 
A similar analysis gives the upper bound 
\begin{align}
\label{eq:upperbound}
\Phi_{i,n} ({d_n^{i,\cE}}) = { \frac{d_n^{i,\cE}} {q_n^i} + c_n \frac{ 1-q_n^i}{2 q_n^i} \left[ \sqrt{ c_n^2 + \frac{ 4 d_n^{i,\cE}} { (1-q_n^i)^2}} +c_n \right] } \; ,
\end{align}
with the same confidence level. 
Then, to satisfy Eq.~\eqref{eq:nphotonerror},
it suffices to set
\begin{align}
\nonumber
c_n^2 ( \epsilon_{\rm DSP})= 4   | \log( \epsilon_{\rm DSP}/24) + n \log (1/2)| \;.
\end{align}

\vspace{2.cm}

To complete the estimation procedure, we
invert Eqs.~\eqref{eq:lowerbound} and \eqref{eq:upperbound}
and obtain
\begin{align}
\label{eq:finalbounds}
& \sum_{n \ge 0}  q_n^i d_n^\cE +c_n ( \epsilon_{\rm DSP})  \sqrt{    q_n^i (1-q_n^i) d_n^\cE} \ge D^{i,\cE} \; , \\
\nonumber
& D^{i,\cE} \ge \sum_{n \ge 0}   q_n^i d_n^\cE - c_n ( \epsilon_{\rm DSP}) \sqrt{   q_n^i (1-q_n^i) d_n^\cE} \; .
\end{align}
We can then execute a  program 
that finds the minimum values of $d_0^\cE$ and $d_1^\cE$
subject to Eqs.~\eqref{eq:finalbounds}.
For instance,  a quadratic program can be used to search
  $\sqrt{q_n^i d_n^{\cE*}}$. Such minimum values
  will be used in Eqs.~\eqref{eq:desiredbound} and \eqref{eq:keyrate} to obtain the key rate.

A technical remark is in order. 
When $n \rightarrow \infty$, $q_n^i (1-q_n^i) \rightarrow 0$
exponentially fast in $n$.
Then, the contribution of large-$n$ terms in Eqs.~\eqref{eq:finalbounds} is negligible.
We can set a suitable cutoff $n_{\max} \ge n$ in the number of photons
per pulse in our analysis, to avoid 
unnecessary computational overheads
in  finding $d_0^{\cE*}$ and $d_1^{\cE*}$, and with an
insignificant impact in the estimated values.

%%%%%%%%%%%%%%%%%%%
\section{Conclusions}
We analyzed general  photon-number splitting
attacks and pointed out that previous security analyses on decoy-state
protocols for QKD made a strong assumption on the attack. We provided 
an estimation procedure that  sets a lower bound on the 
size of the secure, distributed key, with the corresponding confidence levels.
Our procedure requires executing a program to find the minimum values
of the number of detections due to empty and single-photon pulses, subject
to constraints that are determined by the results of the protocol and by the 
desired security parameter.
It results in rigorous security guarantees even if Eve
correlates her attack according to the number of photons in the pulse.

We emphasize that our estimation procedure is not unique: Any time
that a confidence interval can be set as a function
of publicly available information for general attacks, then an estimation procedure is possible. 
In addition, our choice of confidence intervals and 
$\epsilon_n$ is not essential and could be further optimized to improve
the size of the secure key.

\section{Acknowledgments} 
We thank Jane Nordholt, Kevin McCabe, Raymond Newell, Charles Peterson, and  Stephanie Wehner for discussions.
We thank the Laboratory Directed Research and Development (LDRD) Program
at Los Alamos National Laboratory for support.

% \clearpage

%%%%%%%%%%%%%%%%%%%%%%%%%%

\begin{appendix}
\section{Properties of $Z^E_i$}
\label{App:AttackStatistics}

We let $X \in \{0,1,\ldots,K\}$ be a random variable and $f(X)$ the probability distribution.
The random variable $Y \in \{0,1,\ldots,K\}$ has the conditional distribution $g(Y|X)$.
The probability of $Y$ is $h(Y) = \sum_{X=0}^K g(Y|X) f(X)$.
Then, it is easy to show
\begin{align}
\sigma^2_Y = E[\sigma^2_{Y|X} ]+ \sigma^2_{E[Y|X]} \; ,
\end{align}
where 
\begin{align}
\nonumber
\sigma^2_Y =\sum_{Y=0}^K h(Y) Y^2 - \left( \sum_{Y=0}^K h(Y) Y\right)^2 \; 
\end{align}
is the variance of $Y$.
Also,
\begin{align}
\nonumber
E[Y|X] = \sum_{Y=0}^K g(Y|X) Y \; 
\end{align}
is the expected value of $Y$ when given $X$, 
\begin{align}
\nonumber
\sigma^2_{E[Y|X]} = \sum_{X=0}^K f(X) E[Y|X] - \left (\sum_{X=0}^K f(X) E[Y|X] \right)^2 \; 
\end{align}
is the variance of $E[Y|X]$, 
\begin{align}
\nonumber
\sigma^2_{Y|X} = \sum_{Y=0}^K g(Y|X) Y^2 - \left (\sum_{Y=0}^K g(Y|X) Y \right)^2 \; 
\end{align}
is the variance of $Y$ when given $X$, and
\begin{align}
\nonumber
E[\sigma^2_{Y|X} ]= \sum_{X=0}^K f(X) \sigma^2_{Y|X}  \; 
\end{align}
is the expected value of such a variance.

In the attack discussed in Sec.~\ref{sec:attack}, $K$ is fixed and the distribution of $k_n$ satisfies
\begin{align}
\nonumber
E[k_n]=p_n K \; ,\\
\nonumber
\sigma^2_{k_n} = p_n (1-p_n) K \; .
\end{align} 
Next, $d^{\cE}_n$ is chosen such that, when given $k_n$,
\begin{align}
\nonumber
E[d^{\cE}_n|k_n] = y_n k_n \; ,\\
\nonumber
\sigma^2_{d_n^\cE|k_n} = \tau^2 y_n (1-y_n) k_n \; .
\end{align} 
It follows that
\begin{align}
\nonumber
E[\sigma^2_{d_n^\cE|k_n} ] = \tau^2 y_n (1-y_n) p_n K \; , \\
\nonumber
\sigma^2_{E[d_n^\cE|k_n]}= (y_n)^2 \sigma^2_{k_n}= (y_n)^2 p_n (1-p_n) K \; .
\end{align}
Then,
\begin{align}
\nonumber
\sigma^2_{d_n^\cE} = \tau^2 y_n (1-y_n) p_n K + (y_n)^2 p_n (1-p_n) K \; .
\end{align}
When given $d^\cE_n$,  the distribution for $d_n^{i,\cE}$ satisfies
\begin{align}
\nonumber
E[d_n^{i,\cE}|d_n^{\cE}] = q_n^i d_n^{\cE} \; , \\
\nonumber
\sigma^2_{d_n^{i,\cE}|d_n^{\cE}} = q_n^i (1-q_n^i) d_n^{\cE} \; .
\end{align}
Therefore,
\begin{align}
\nonumber
\sigma^2_{E[d_n^{i,\cE}|d_n^{\cE}]} = (q_n^i)^2 \sigma^2_{d_n^{\cE}} \; ,
 \\
 \nonumber
E[\sigma^2_{d_n^{i,\cE}|d_n^{\cE}}]= q_n^i (1-q_n^i) y_n p_n K \; .
\end{align}
Also,
\begin{align}
\nonumber
\sigma^2_{d_n^{i,\cE}} & =(q_n^i)^2 \sigma^2_{d_n^{\cE}} + q_n^i (1-q_n^i) y_n p_n K \\
\nonumber
&= (q_n^i)^2 [ \tau^2 y_n (1-y_n) p_n K + (y_n)^2 p_n (1-p_n) K] + \\
\nonumber & + q_n^i (1-q_n^i) y_n p_n K \\
\nonumber
& =[ (\tau^2-1) q_n^i (1-y_n) + (1-q_n^i y_n p_n)] q_n^i y_n p_n K \; .
\end{align}
The first term on the {\em rhs} vanishes when $\tau=1$.
The second term is
\begin{align}
\nonumber
(1-q_n^i y_n p_n) q_n^i y_n p_n K = (1-q^i y_n p^{\mu^i}_n) q^i y_n p^{\mu^i}_n K\; ,
\end{align}
 so that
\begin{align}
\nonumber
\sum_{n \ge 0} \sigma^2_{d_n^{i,\cE}} =  q^i Y(\mu^i) K 
- (q^i)^2 K \sum_{n \ge 0} [y_n p_n^{\mu^i} ]^2 \; ,
\end{align}
for $\tau=1$.
Moreover, since $\sum_{n \ge 0} [y_n p_n^{\mu^i} ]^2 \ll Y(\mu^i) \ll 1$, 
then
\begin{align}
\nonumber
\sum_{n \ge 0} \sigma^2_{d_n^{i,\cE}} \approx  q^i Y(\mu^i) (1-Y(\mu^i) )K  \; ,
\end{align}
which shows that the case discussed in Sec.~\ref{sec:decoys}, i.e. the i.i.d. assumption, 
corresponds to choosing $\tau=1$ in this case.

%%%%%%%%%%%%%%%%%%%%%%%%%%%%%%%%%%%%%%%%%%%%%%%

\section{Chernoff bound}
\label{app:chernoff}
Chernoff's bound~\cite{Ho63} sets a bound on the probabilities of ``rare'' events
as a function of the standard deviation of the corresponding distribution. 
More precisely, we let $X_1,X_2,\dots ,X_n$ be a set of i.i.d. random variables that satisfy $|X_j| \le 1$
and define $X = \sum_j X_j$. A general version of Chernoff's bound implies
\begin{align}
\label{eq:chernofferror0}
\Pr[X >  E[X] + c \sigma] \le  \exp \{- c^2/4 \} \; ,
\end{align}
where $\sigma=n^{1/2} \sqrt{E[(X_j)^2] -( E[X_j])^2}$ is the standard deviation.
 For the special case of the binomial distribution where $X_j=1$ with probability $a$
 and $X_j=0$ otherwise,
\begin{align}
\nonumber
\sigma &= \sqrt{n a(1-a) } \; , \\
\nonumber
E[X] & = n a\; ,
\end{align}
and
\begin{align}
\nonumber
\Pr [X> k] &= I_{a} (k,n-k+1) \\
\label{eq:chernoffboundBINOMIAL}
& \le  \exp \{- (k - n a )^2/(4 a (1-a) n) \} \; .
\end{align}
Here, $I_{a} (k,n-k+1) $ is the so-called regularized incomplete beta function.
To satisfy $\Pr [X> k]  \le \epsilon$, it suffices to choose $c$ such that
\begin{align}
\label{eq:chernofferror}
|c| = 2 \sqrt{ |\log \epsilon|} \; .
\end{align}

%%%%%%%%%%%%%%%%%%%%%%%%%%%%%%%%%%%%%%%%%%%%%%%
\section{Calculations of errors}
\label{app:errors}

If $\epsilon_n \le (\epsilon/12)(1/2)^n$, then
\begin{align}
\label{eq:firsterror}
\bar \epsilon & = \sum_{i}\sum_n  \epsilon_n \\
\nonumber 
& \le 3 (\epsilon/12) .2 = \epsilon/2 \; ,
\end{align}
where we considered that three sources $i$ are involved in the DSP.

Chernoff's bound for the binomial distribution [Eq.~\eqref{eq:chernoffboundBINOMIAL}]
implies that 
\begin{align}
\nonumber
\epsilon_n & \le  2  \exp \left \{ - \frac {(u_n^i - q_n^i d_n^\cE )^2} {4 q_n^i (1-q_n^i) d_n^\cE}  \right \} \; .
\end{align}
If we set,
\begin{align}
\label{eq:maxDU}
u_n^i=\phi_{i,n}^{-1}(d_n^\cE) =  q_n^i d_n^\cE + c_n \sqrt{     q_n^i (1-q_n^i) d_n^\cE} \; ,
\end{align}
then 
\begin{align}
\nonumber
\epsilon_n & \le  2 \exp\{-c_n^2/4\}\; ,
\end{align}
as in Eq.~\eqref{eq:errorbound3}.
Replacing $u_n^i$ by $d_n^{i,\cE}$ and $d_n^\cE$
by $\phi_{i,n}(d_n^{i,\cE})$ in Eq.~\eqref{eq:maxDU},
and solving the resulting quadratic equation,
we obtain
\begin{align}
\nonumber
&\sqrt{\phi_{i,n}(d_n^{i,\cE})} = \left[ - c_n  \sqrt{    q_n^i (1-q_n^i)} +\right. \\ 
\nonumber
& \left. + \sqrt{c_n^2   q_n^i (1-q_n^i) + 4 q_n^i d_n^{i,\cE}} \right] / (2q_n^i) \; .
\end{align}
That is,
\begin{align}
\nonumber
\phi_{i,n}(d_n^{i,\cE}) &= \frac{d_n^{i,\cE}}{q_n^i} + \frac{c_n^2 (1-q_n^i)}{2 q_n^i} - \\
\nonumber
&- \frac{c_n \sqrt{c_n^2 (1-q_n^i) [(1-q_n^i) + 4 d_n^{i,\cE}] } }{2 q_n^i} \; ,
\end{align}
that yields Eq.~\eqref{eq:lowerbound}.
Changing $c_n \rightarrow -c_n$ provides the upper bound without changing $\epsilon_n$,
i.e., the confidence level.

\end{appendix}


\begin{thebibliography}{10}

\bibitem{Wie83}
S. Wiesner. Conjugate coding. Sigact News, 15(1):78 -- 88, 1983.

\bibitem{BB84}
C. H. Bennett and G. Brassard. Quantum cryptography:
Public-key distribution and coin tossing. In Proceedings of
IEEE International Conference on Computers, Systems and
Signal Processing, pages 175--179, 1984.

\bibitem{Eke91}
A. K. Ekert. Quantum cryptography based on Bell's theorem.
Phys. Rev. Lett., 67:661, 1991.

\bibitem{Ben92}
C. H. Bennett. Quantum cryptography using any two
nonorthogonal states. Phys. Rev. Lett., 68(21):3121--3124, 1992.

\bibitem{May01}
D. Mayers. Quantum key distribution and string oblivious
transfer in noisy channels. In Advances in Cryptology --
CRYPTO `96, volume 1109 of Lecture Notes in Computer Science,
pages 343--357. Springer, 1996.

\bibitem{SP00}
P. Shor and J. Preskill. Simple proof of security of the BB84
quantum key distribution protocol. Phys. Rev. Lett., 85:441--444,
2000.

\bibitem{Ren05}
Renato Renner. Security of Quantum Key Distribution. Swiss Federal
Institute of Technology Zurich. PhD Thesis. Zurich, 2005.

\bibitem{Zur82}
W. Zurek. A Single Quantum Cannot be Cloned. 
Nature 299, 802--803, 1981.

\bibitem{BBBSS92}
C.H. Bennett, F. Bessette, G. Brassard, L. Salvail, and J. Smolin.
Esperimental quantum cryptography.
Journal of Cryptology, 5: 3--28, 1992.

\bibitem{SWF07}
Tobias Schmitt-Manderbach, et.al. Experimental Demonstration of Free-Space
Decoy-State Quantum Key Distribution over 144Km. Phys. Rev. Lett., 98: 010504, 2007.

\bibitem{RPH09}
D. Rosenberg, et.al. Practical long-distance quantum key distribution system using decoy levels.
New J. Phys., 11: 045009, 2009.


\bibitem{Lut00}
Norbert L\"utkenhaus. Security against individual attacks for realistic quantum key distribution. Phys. Rev. A, 61: 052304, 2000.

\bibitem{LJ02}
Norbert L\"utkenhaus and Mika Jahma.
Quantum key distribution with realistic states:
photon-number statistics in the photon-number splitting
attack. New Journal of Physics, 4: 1--44, 2002.

\bibitem{GLLP04}
D. Gottesman, H.-K. Lo, N. L\"utkenhaus, and 
J. Preskill. Quantum key distribution with imperfect devices. Quantum Information and Computation,
4: 325, 2004.

\bibitem{Hwa03}
Won-Young Hwang. Quantum Key Distribution with High Loss: Toward Global Secure Communication.
Phys. Rev. Lett., 91: 057901, 2003.

\bibitem{LMC05}
Hoi-Kwong Lo, Xionfeng Ma, and Kai Chen. 
Decoy State Quantum Key Distribution.
Phys. Rev. Lett., 94: 230504 (2005).

\bibitem{RH09}
Patrick Rice and Jim Harrington.
Numerical analysis of decoy state quantum key distribution protocols.
E-print arXiv:0901.0013, 2009.



\bibitem{Note3}
 Our definition of $D^{i,\cE}$ does
not imply that the same $n$-photon pulse created by Alice is detected by Bob, as Eve may be replacing each pulse
at her will. $D^{i,\cE}$ only gives information about detections corresponding to those clock cycles
in which Alice prepared the pulse with source $i$. Security will follow from the analysis on the bit error
rates that will subtract those detections corrupted by Eve as implied by Eq.~\eqref{eq:keyrate}.


\bibitem{Ho63}
Wassily Hoeffding. Probability Inequalities for Sums of Bounded Random Variables. J. Amer. Statist. Assoc.,  58 (301): 13, 1963. H. Chernoff. A Note on an Inequality Involving the Normal Distribution. Annals of Probability,
 9: 533, 1981.


\bibitem{Note1}
The exact values of $Y(\mu^i)$ are irrelevant
as Alice and Bob cannot have exact estimates of the channel parameters
such as $\eta$, in general. If Eve changes these values slightly,
such changes will not be noticed.

%\bibitem{Note2}
%A simple way to understand this is the following. Suppose there is a box that turns on
%whenever source $i$ is used at least once. That box not only turns on but also tells
%the value of $x^i$ and then the value of the number of detections $D^i$ by Bob.
%Whenever the box does not turn on, we do not count those events for the random variable
%$Z^E_i$.

%\bibitem{BE}
%Berry, Andrew C.. The Accuracy of the Gaussian Approximation to the Sum of Independent Variates. Transactions of the American Mathematical Society 49 (1): 122--136, 1941.
%Esseen, Carl-Gustav. A moment inequality with an application to the central limit theorem. Skand. Aktuarietidskr. 39: 160--170,1956.


\end{thebibliography}
\end{document}